\documentclass[runningheads]{llncs}
\usepackage{amsmath,amsfonts}
\usepackage{algorithmic}
\usepackage{algorithm}
\usepackage{array}
\usepackage{mathtools}
\usepackage{graphicx}
\usepackage{textcomp}
\usepackage{adjustbox}
\usepackage{multirow}
\usepackage{makecell}
 \usepackage{caption}
\usepackage{textcomp}
\usepackage{stfloats}
\usepackage{url}
\usepackage{verbatim}
\usepackage{cite}
\usepackage{float}
\usepackage{subfig}
\usepackage{tikz}
\usepackage{calc}
\usepackage{bm}
\usepackage{gensymb}
\usepackage{booktabs}
\usepackage{amssymb}
\usepackage{pifont}
\usepackage{xcolor}
\definecolor{citecolor}{HTML}{2980b9}
\definecolor{linkcolor}{HTML}{c0392b}

\newcommand\blfootnote[1]{%
  \begingroup
  \renewcommand\thefootnote{}\footnote{#1}%
  \addtocounter{footnote}{-1}%
  \endgroup
}

\newcommand{\cmark}{\ding{51}}%
\newcommand{\xmark}{\ding{55}}%

\newcommand{\etal}{\textit{et al}.}
\newcommand{\ie}{\textit{i}.\textit{e}.}
\newcommand{\eg}{\textit{e}.\textit{g}.}

\makeatletter
\newcommand\figcaption{\def\@captype{figure}\caption}
\newcommand\tabcaption{\def\@captype{table}\caption}
\makeatother

\usepackage[hidelinks,pagebackref=true,breaklinks=true,colorlinks,bookmarks=false,citecolor=citecolor,letterpaper=true,linkcolor=linkcolor]{hyperref}
\usepackage{cleveref}

\begin{document}
\title{Steerable Pyramid Transform Enables \\Robust Left Ventricle Quantification}

\author{Xiangyang Zhu\inst{1} \and
Kede Ma\inst{1}$^\dagger$ \and
Wufeng Xue\inst{2}$^\dagger$
}
\institute{City University of Hong Kong \and Shenzhen University}

\maketitle 
\blfootnote{$\dagger$ Corresponding authors.}

\begin{abstract}
Predicting cardiac indices has long been a focal point in the medical imaging community. While various deep learning models have demonstrated success in quantifying cardiac indices, they remain susceptible to mild input perturbations, \eg, spatial transformations, image distortions, and adversarial attacks. This vulnerability undermines confidence in using learning-based automated systems for diagnosing cardiovascular diseases. In this work, we describe a simple yet effective method to learn robust models for left ventricle (LV) quantification, encompassing cavity and myocardium areas, directional dimensions, and regional wall thicknesses. Our success hinges on employing the biologically inspired steerable pyramid transform (SPT) for fixed front-end processing, which offers three main benefits. First, the basis functions of SPT align with the anatomical structure of LV and the geometric features of the measured indices. Second, SPT facilitates weight sharing across different orientations as a form of parameter regularization and naturally captures the scale variations of LV. Third, the residual highpass subband can be conveniently discarded, promoting robust feature learning. Extensive experiments on the Cardiac-Dig benchmark show that our SPT-augmented model not only achieves reasonable prediction accuracy compared to state-of-the-art methods, but also exhibits significantly improved robustness against input perturbations. Code is available at \url{https://github.com/yangyangyang127/RobustLV}.

\end{abstract}
\section{Introduction}
\label{sec:introduction}
\begin{figure*}[!t]
\centerline{\includegraphics[width=12.0cm]{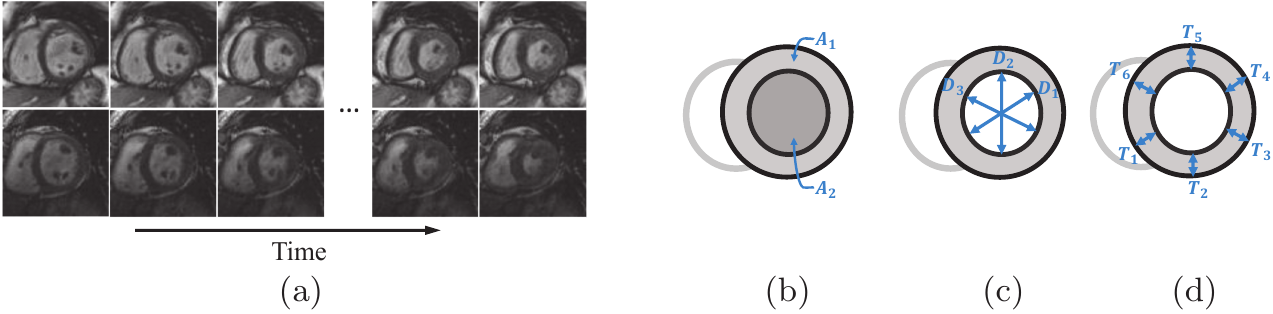}}
\caption{Adopted LV quantification setting. (a) Two example sequences of CMR slices with different shapes, brightness, and contrast. (b) LV myocardium area ($A_1$) and cavity area ($A_2$). (c) Three directional dimensions of the cavity ($D_{1} \sim D_{3}$). (d) Six regional wall thicknesses ($T_{1} \sim T_{6}$).}
\label{fig:LV_indices}
\end{figure*}

In the academic and clinical communities, quantifying physiological cardiac indices, particularly those pertaining to the left ventricle (LV), has garnered significant attention~\cite{Peng2016Areview}.
Cardiovascular magnetic resonance (CMR) based LV quantification aims to provide a comprehensive, objective, and precise evaluation of LV pathological changes by predicting structural indices. In this work, we adopt a commonly-used setting for LV quantification, which takes a sequence of CMR slices\footnote{More specifically, the input comprises a sequence of CMR slices from the mid-cavity in the short-axis view, obtained using the cine CMR protocol.} acquired within one cardiac cycle as input and predicts a set of structural indices for each slice, \ie, cavity and myocardium areas, three directional dimensions, and six regional wall thicknesses \cite{xue2017direct, xue2018full}, totaling eleven indices as shown in Fig. \ref{fig:LV_indices}. 

Existing approaches to LV quantification focus primarily on improving prediction accuracy, often measured in terms of mean absolute error (MAE) between model predictions and human annotations.
This has led to an impressive series of sophisticated methods, ranging from CMR statistics-based algorithms to deep learning-based methods~\cite{Zhen2014Direct, afshin2013regional, xue2017direct, xue2018full}. Despite the demonstrated successes, there has been limited research
into testing and improving the robustness of LV quantification methods. Not surprisingly, even minor input perturbations---arising from phases like acquisition, storage, transmission, and clinical applications---can significantly compromise the performance of nearly all existing LV quantification methods. This vulnerability impedes reliable diagnosis and effective risk stratification for multifarious cardiovascular diseases \cite{Ma2021Understanding}. 

Instead of focusing solely on achieving state-of-the-art quantification performance, we switch our attention toward testing and improving the robustness of LV quantification. Central to our approach is the use of the steerable pyramid transform (SPT) \cite{Simoncelli1995Thesteerable} as a fixed biologically inspired front-end. Despite its simplicity, SPT offers three key advantages. First, the convolutional basis filters of SPT are localized in orientation and scale, providing translation-equivariant (\ie, aliasing-free) and rotation-equivariant (\ie, steerable) responses. This property aligns well with the anatomical structures and spatiotemporal dynamics of LV, as well as the geometric characteristics of the estimated indices. Second, SPT facilitates straightforward implementation of parameter sharing across orientations as a form of regularization, and explicitly models the scale variations of LV throughout a cardiac cycle. Third, the residual highpass subband, identified as a potential source of model vulnerability \cite{Wang2020High}, can be conveniently discarded to promote more robust cardiac representation learning. Experiments demonstrate that our SPT-augmented method not only well predicts the eleven LV indices but also substantially enhances robustness against existing methods when exposed to spatial transformations, image distortions, and adversarial attacks.

\section{Steerable Pyramid Transform}\label{sec:rw}

\begin{figure*}[t]
\begin{minipage}[c]{0.5\textwidth}
\centering
\includegraphics[width=0.99\textwidth]{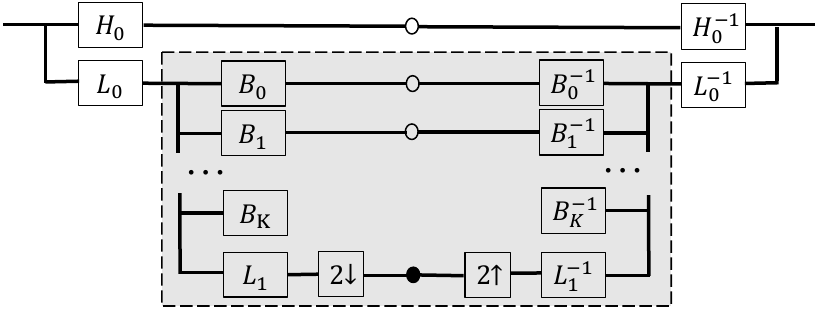}
\caption{The block diagram of SPT (including both analysis and synthesis transforms), reproduced from \cite{Simoncelli1995Thesteerable}. The pyramid is recursively constructed by inserting a copy of the shaded portion of the diagram at the location of the solid circle \cite{Simoncelli1995Thesteerable}.}
\label{fig:SPT_framework}
\end{minipage}\hspace{12pt}
\begin{minipage}[c]{0.45\textwidth}
\centering
\centerline{\includegraphics[width=0.82\textwidth]{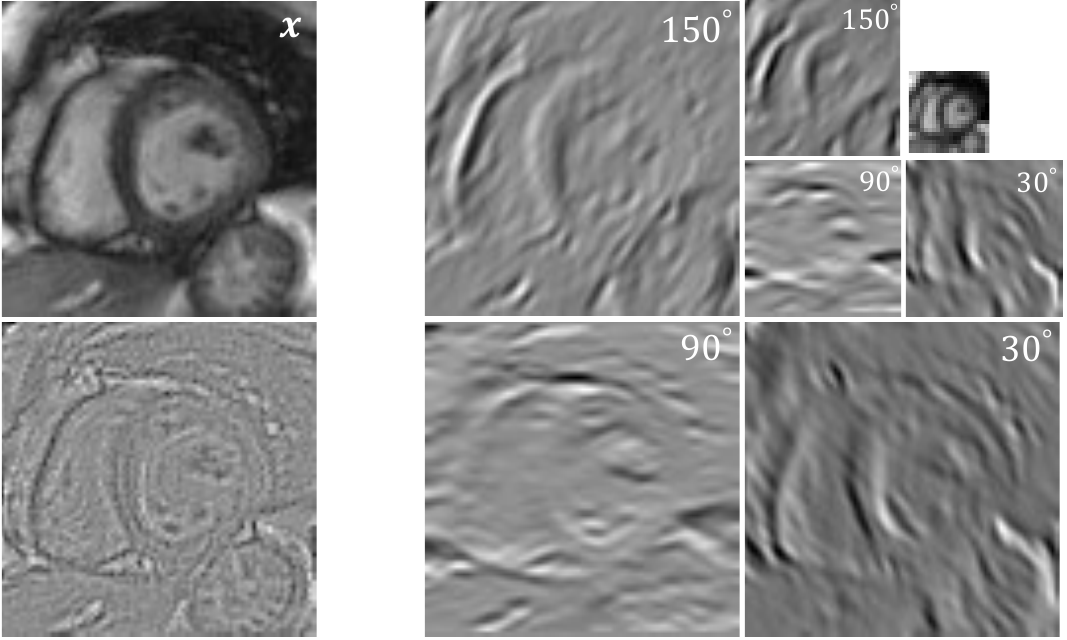}}
\caption{SPT of a CMR image. Left: CMR image $\bm x$ and its residual highpass subband. Right: Subbands oriented at $30\degree$, $90\degree$, and $150\degree$ and the residual lowpass subband.}
\label{fig:SPT_decompose}
\end{minipage}
\end{figure*}

SPT is a linear multi-scale multi-orientation image decomposition, developed by Simoncelli \etal~\cite{Simoncelli1995Thesteerable, simoncelli1992shiftable} in the 1990s to address the limitations of orthogonal wavelets. The basis functions are $K$-th order directional derivative filters, which come in varying sizes and orientations. As shown in Fig. \ref{fig:SPT_framework}, SPT begins by decomposing an image $\bm x$ into high and lowpass subbands using filters $H_0$ and $L_0$. The lowpass subband is then separated into a set of oriented subbands by a bank of filters $\{B_k\}_{k=0}^K$ and a lower-pass subband by the filter $L_1$ with a stride of two (\ie, down-sampling by a factor of two). This lower-pass subband can be recursively decomposed to build the steerable pyramid.
An example decomposition of a CMR image, containing three orientations at two scales, is shown in Fig. \ref{fig:SPT_decompose}. 

Over the past three decades, the use of SPT has proven beneficial in a variety of image processing and computer vision applications, spanning from low-level tasks such as image denoising \cite{Portilla2003Image}, image compression \cite{buccigrossi1999image}, image steganalysis \cite{lyu2006steganalysis} and image quality assessment \cite{sheikh2006image}, to middle-level tasks like texture synthesis \cite{Portilla2000Aparametric} and image segmentation \cite{chen2005adaptive}, and to high-level tasks including object detection \cite{torralba2003context} and visual tracking \cite{jepson2003robust}. In the era of deep learning, SPT continues to find innovative uses. Meyer \etal~\cite{meyer2018phasenet} proposed PhaseNet for video frame interpolation. Bhardwaj \etal~\cite{Bhardwaj2021Unsupervised} presented an unsupervised information-theoretic perceptual quality metric using SPT as the fixed input transform. Deng \etal~\cite{Deng2020Mimamonet} adopted inter-frame phase differences by SPT rather than optical flow to enhance video emotion recognition. Parthasarathy and Simoncelli relied on SPT to mimic the V1 complex cells~\cite{parthasarathy2020self}. While most existing work leverages SPT to improve computational prediction accuracy, our research explores the added robustness of SPT for LV quantification.

\section{Proposed Method}\label{sec:method}
\begin{figure*}[!t]
\centerline{\includegraphics[width=12.2cm]{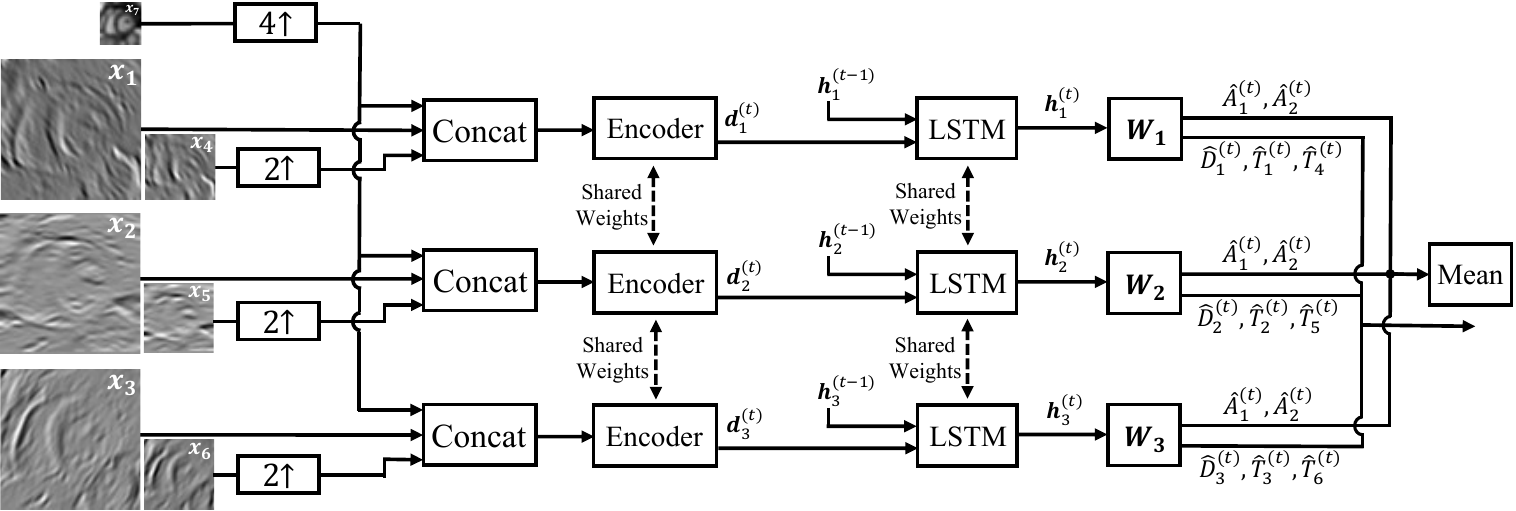}}
\caption{The system diagram of our SPT-augmented method for LV quantification.}
\label{fig:whole_framework}
\end{figure*}

In this section, we describe our SPT-augmented method for achieving robust and accurate LV quantification. At a high level, we decompose each slice in a CMR image sequence into a steerable pyramid with three orientations at two scales. 
We then use a five-layer CNN to map the pyramid to three oriented feature vectors, which are fed to
a shared long short-term memory (LSTM)~\cite{hochreiter1997long} to predict the
eleven structural LV indices. Learning 
is driven by the MAE and a multi-task regularizer. The system diagram is shown in Fig.~\ref{fig:whole_framework}.

\subsection{SPT Decomposition}
\begin{figure*}[t]
\hspace{-20pt}
\begin{minipage}[c]{0.5\textwidth}
\centering
\includegraphics[width=0.9\textwidth]{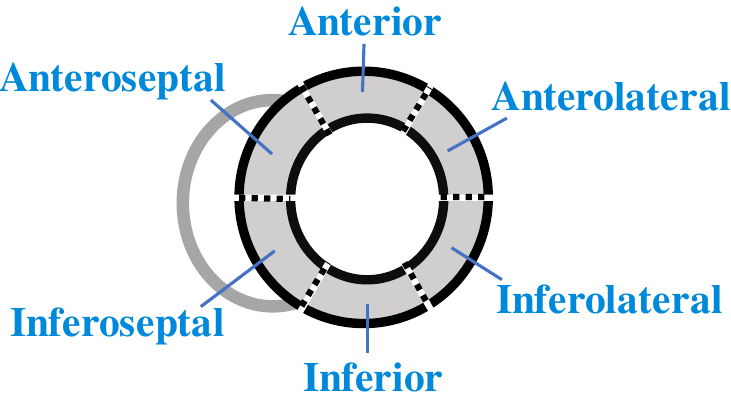}
\vspace{4pt}
\\
\includegraphics[width=0.4\textwidth]{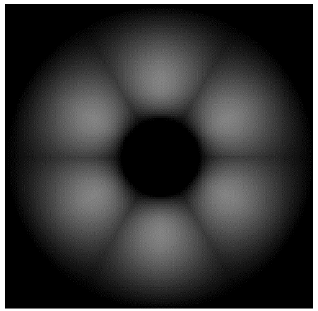}
\caption{Top: Regional walls of the mid-cavity LV myocardium by the $17$-segment model \cite{Cerqueira2002Standardized}. Bottom: the corresponding subbands of SPT in the Fourier domain.}
\label{fig:direction_matching_of_SPT}
\end{minipage}\hspace{12pt}
\begin{minipage}[c]{0.55\textwidth}
\centering
\begin{tikzpicture}
\fontsize{6pt}{1pt}\selectfont
\node [matrix, inner sep=0pt, column sep=2pt, row sep=1ex,
layer/.style={rectangle, draw, fill=white, rotate=90, minimum width=40ex, inner ysep=3.3pt},
numpars/.style={rotate=90, font=\scriptsize, left}] (analysis) {
\node[layer] {convolution $|\; 7 \;|\;3 \times 60 \;|\;3$}; &
\node[layer] {maxpooling $|\; 2 \;|\;2 $}; &
\node[layer] {convolution $|\; 5 \;|\;60\times 120 \;|\;2$}; &
\node[layer] {maxpooling $|\; 2 \;|\;2 $}; &
\node[layer] {convolution $|\; 5 \;|\;120\times 240 \;|\;2$}; &
\node[layer] {maxpooling $|\; 2 \;|\;2 $}; &
\node[layer] {convolution $|\; 3\;|\;240\times 480 \;|\;1$}; &
\node[layer] {maxpooling $|\; 2 \;|\;2 $}; &
\node[layer] {convolution $|\; 3\;|\;480\times 480 \;|\;1$}; &
\node[layer] {$\ell_2$-pooling $|\; 5 \;|\;5 $}; &
\node[layer] {fc $|\;480\times 100$}; &\\

\node[numpars] {9000}; &
\node[numpars] {0}; &
\node[numpars] {180360}; &
\node[numpars] {0}; &
\node[numpars] {720720}; &
\node[numpars] {0}; &
\node[numpars] {1038240}; &
\node[numpars] {0}; &
\node[numpars] {2075040}; &
\node[numpars] {0}; &
\node[numpars] {48100}; &
\\
};
\end{tikzpicture}
\caption{The architecture of the CNN. The parameterization of the convolution and pooling layers is denoted as ``kernel size $\mid$ input channel $\times$ output channel $\mid$ stride" and ``kernel size $\mid$ stride", respectively. The number of parameters for each layer is given at the bottom, yielding a total of $4.07$ million.}
\label{fig:cnn_structure}
\end{minipage}
\end{figure*}

Given an input slice $\bm x \in \mathbb{R}^{M\times N}$, we decompose it into a steerable pyramid $\{\bm x_i\}_{i=0}^7$, where $\bm x_0 \in \mathbb{R}^{M\times N}$ and $\bm x_7 \in \mathbb{R}^{\left\lfloor \frac{M}{4}\right\rfloor \times \left\lfloor\frac{N}{4}\right\rfloor}$ are residual highpass and lowpass subbands, respectively. $\bm x_1, \bm x_2, \bm x_3 \in \mathbb{R}^{M\times N}$ are three oriented subbands to $30\degree$, $90\degree$, and $150\degree$, respectively, at Scale 1 (\ie, the same scale as the original slice). Similarly, $\bm x_4, \bm x_5, \bm x_6 \in \mathbb{R}^{\left\lfloor \frac{M}{2}\right\rfloor \times \left\lfloor\frac{N}{2}\right\rfloor}$ are three oriented subbands at Scale 2. We use two scales because the LV slice $\bm x$ is often a small region in the cardiac image, which is of limited resolution (\eg, $80\times 80$). We use three orientations at $\{30\degree, 90\degree, 150\degree\}$ because the mid-cavity myocardium approximates a 2D torus in the short-axis view, and both directional dimensions and regional wall thicknesses are distributed along these orientations, as illustrated in Figs. \ref{fig:LV_indices} and \ref{fig:direction_matching_of_SPT}.

After establishing the hyperparameter setting of the steerable pyramid, the next step is to organize the pyramid as the input to the encoder. To achieve this, we make two key observations. First, from the computational neuroscience literature \cite{awasthi2011faster, bar2004visual}, humans rely primarily on low-frequency information for object recognition. In contrast, convolutional neural networks (CNNs) tend to exploit high-frequency information that is imperceptible to the human eye for decision-making \cite{Wang2020High, Yin2019AFourier}, 
resulting in generalization behaviors that deviate perceptually from humans. This motivates us to discard the highpass subband $\bm x_0$ as a way of forcing the CNN to learn robust cardiac representations from lowpass and oriented subbands. Second, SPT allows for independent representations of orientation and scale \cite{Simoncelli1995Thesteerable}. In the context of LV quantification, orientation is more pertinent to spatial predictions of LV indices, such as the three directional dimensions and six regional wall thicknesses. Scale, on the other hand, is more relevant to temporal predictions of these indices, which vary with myocardial contraction and relaxation. Thus, we gather all scale subbands of the same orientation, along with the lowpass subband, as the input to the encoder. The parameters of the encoder are shared across different orientations. This data organization can be viewed as a form of parameter regularization, encouraging the encoder to attend to different directional visual cues. Additionally, we model the scale variations of LV from diastole to systole and across different subjects as the input subbands span three different scales. An efficient implementation of the proposed data organization is to allocate orientation and scale/lowpass subbands in the batch and channel dimensions, respectively. Bilinear interpolation is used to upsample small-scale subbands to the size of the original image.

\subsection{LV Quantification}
We adopt a generic five-layer CNN to extract spatially oriented features. Within the CNN, each layer applies a bank of convolution filters, whose responses are batch-normalized~\cite{ioffe2015batch} and half-wave rectified~\cite{nair2010rectified}. Maxpooling is added to the first four convolution layers
with a stride of two. We compensate for spatial downsampling by doubling the number of channels (\ie, filters).
We use global $\ell_2$-pooling to summarize spatial
information into a fixed-length feature vector, which is further processed by a fully connected layer with dropout regularization~\cite{srivastava2014dropout}. After the CNN, we represent the  LV slice $\bm x$ by $\{\bm d_i\}_{i=1}^3$, where $\bm d_i\in  \mathbb{R}^{N_{c}\times 1}$ is the $i$-th orientation feature vector. The detailed specification is shown in Fig.~\ref{fig:cnn_structure}.

Next, we adopt an LSTM to model the temporal dynamics of LV. The LSTM cell relies on the hidden state vector (also known as the output vector of the LSTM) of the previous LV slice, $\bm h^{(t-1)}\in \mathbb{R}^{N_c\times 1}$ and the feature vector of the current slice, $\bm d^{(t)}$ to compute three gates, $\bm f^{(t)}, \bm i^{(t)}, \bm o^{(t)}\in\mathbb{R}^{N_c \times 1} $ and one intermediate cell state vector $\tilde{\bm c}^{(t)} $.
$\bm f^{(t)}$, $\bm i^{(t)}$, and $\tilde{\bm c}^{(t)}$ together are used to update the cell state vector:
\begin{align}
    \bm c^{(t)} = \bm f^{(t)}\odot\bm c^{(t-1)} + \bm i^{(t)} \odot \tilde{\bm c}^{(t)},
\end{align}
where $\odot$ denotes the element-wise product. The current hidden state vector is computed by 
\begin{align}\label{eq:ht}
    \bm h^{(t)} = \bm o^{(t)}\odot\mathrm{tanh}(\bm c^{(t)}).
\end{align}
Here, we omit the subscript index $i$ because the LSTM cell is shared across orientations. After the LSTM, we represent the $t$-th LV slice $\bm x^{(t)}$ by $\{\bm h^{(t)}_i\}_{i=1}^3$, based on which we use linear regression to predict the eleven LV indices:
\begin{align}
\label{eq_relation1}
[\hat{A}^{(t)}_{1,i}; \hat{A}^{(t)}_{2,i}; \hat{D}^{(t)}_i; \hat{T}^{(t)}_i; \hat{T}^{(t)}_{i+3}] = \bm W_{i} \bm h_{i}^{(t)} + \bm b_{i},\quad i=1,2,3,
\end{align}
where $\bm W_i\in\mathbb{R}^{5\times N_{c}}$ and $\bm b_i \in \mathbb{R}^{5\times 1}$ are the $i$-th weight matrix and bias term. That is, the $i$-th oriented representations are responsible for estimating two cavity areas $\hat{A}^{(t)}_{1,i}$ and $\hat{A}^{(t)}_{2,i}$, one corresponding directional dimension $\hat{D}_i^{(t)}$, and two corresponding wall thicknesses $\hat{T}_i^{(t)}$ and $\hat{T}_{i+3}^{(t)}$. The final cavity and myocardium areas are computed by averaging predictions from the three orientations:
\begin{align}
    \hat{A}^{(t)}_r = \frac{1}{3}\sum_{i=1}^3\hat{A}^{(t)}_{r,i},\quad r = 1,2. 
\end{align}

\subsection{Loss Function}
For notation convenience, we collectively denote the eleven predictions of the LV indices by $\hat{\bm y} = \bm g_{\bm \theta}(\bm x) \in \mathbb{R}^{11\times 1}$, where $\bm g$ denotes the complete LV quantification method, parameterized by a vector $\bm \theta$. 
We adopt the MAE as the loss function:
\begin{align}
\label{eq_basic_loss}
\ell_1(\bm{\theta};\mathcal{B}) = \frac{1}{\vert\mathcal{B}\vert}\sum_{(\bm x, \bm y)\in\mathcal{B}} \Vert \bm y - \bm g_{\bm \theta}({\bm x}) \Vert_1,
\end{align}
where $\mathcal{B}$ is a minibatch of training examples sampled from the full dataset. To enable a fair comparison with \cite{xue2018full},
we also impose orientation, index, and weight correlation constraints on the final linear regression parameters $\bm W_\mathrm{lr} \coloneqq \{\bm W_1, \bm W_2, \bm W_3\} \in \mathbb{R}^{N_o\times N_i\times N_c}$, where $N_o=3$, $N_i=5$, and $N_c=100$. We assume a tensor Normal distribution \cite{Ohlson2013The} for the parameters:
\begin{align*}
p\left(\bm W_\mathrm{lr} \right)  &= \mathcal{TN}\left(\bm W_\mathrm{lr};\bm{0}, \bm{\Sigma}_o, \bm{\Sigma}_i, \bm{\Sigma}_c \right)
= \mathcal{N}\left(\mathrm{vec}(\bm W_\mathrm{lr});\bm{0}, \bm{\Sigma}_
\mathrm{lr} \right)
= {\left( {2\pi } \right)^{ - N_oN_iN_c/2}} \\ \times & {\prod_{k\in\{o,i,c\}}{{{\left\vert {{\bm{\Sigma}_k}} \right\vert}^{ - N_{o}N_{i}N_{c}/\left( {2{N_k}} \right)}}} } \times \exp \left( { - \frac{1}{2}{{ { {\mathrm{vec}}\left( \bm{W}_\mathrm{lr} \right) } }^{\mathsf{T}}}\bm{\Sigma} _\mathrm{lr}^{ - 1}{\mathrm{vec}}\left( \bm{W}_\mathrm{lr} \right)} \right).
\end{align*}
$\mathrm{vec}(\cdot)$ is the vectorized operation and ${\bm{\Sigma}_\mathrm{lr}} = {\bm{\Sigma}_o} \otimes  {\bm{\Sigma}_i}  \otimes {\bm{\Sigma}_c}$ denotes the Kronecker product of the orientation covariance matrix ${\bm{\Sigma}_o}\in \mathbb{R}^{N_o\times N_o}$, the index covariance matrix ${\bm{\Sigma}_i} \in \mathbb{R}^{N_i\times N_i}$, and the weight covariance matrix ${\bm{\Sigma}_c}\in\mathbb{R}^{N_c\times N_c}$. We further assume that $\bm y_i$ is Laplace distributed with the mean vector estimated by $\hat{\bm y}_i$ and the covariance matrix fixed to identity. As a result, maximizing the posterior distribution of $\bm \theta$ is equivalent to minimizing
\begin{align*}
\ell(\bm \theta, {\bm \Sigma}_\mathrm{lr};\mathcal{B}) = \ell_1(\bm \theta;\mathcal{B})+\lambda \Big( \mathrm{vec}(\bm W_\mathrm{lr})^\mathsf{T} \bm{\Sigma}_\mathrm{lr}^{-1} \mathrm{vec}(\bm W_\mathrm{lr})
- \sum_{k\in\{o,i,c\}}\frac{N_o N_i N_c}{N_k} \ln \vert{{\bm \Sigma}_k} \vert\Big),
\end{align*}
where we assume the remaining parameters $\bm \theta\setminus \bm W_\mathrm{lr}$ are uniformly distributed, whose log-likelihoods are constant and thus can be omitted. The added regularization has two effects. The first term inside the parentheses minimizes the magnitude of the weights $\bm W_\mathrm{lr}$, functioning similarly to $\ell_2$-regularization. The second term maximizes the absolute value of the determinant of $\bm{\Sigma}_k$, thereby encouraging learning of independent features.

\subsection{Robustness Evaluation}\label{subsec:robust_eval} 
\begin{figure}[t]
\centering
\includegraphics[width=1.0\textwidth]{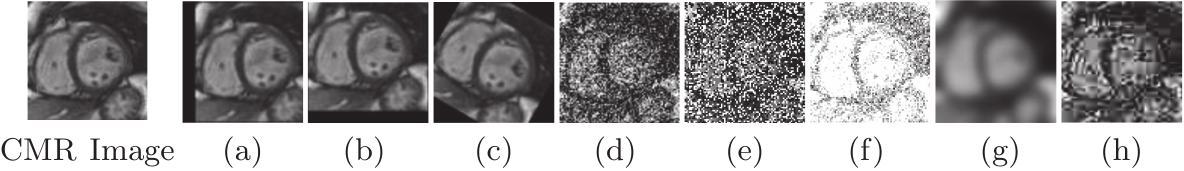}
\caption{Illustration of input perturbations. (a) Horizontal translation. (b) Vertical translation. (c) Rotation. (d) Gaussian noise. (e) Impulsive noise. (f) Rician noise. (g) Gaussian blur. (h) JPEG compression. We do not show examples of adversarial attacks as they are generally imperceptible to the human eye.}
\label{fig:distortions}
\end{figure}

We examine three types of input perturbations: spatial transformations, image distortions, and adversarial attacks. Examples are given in Fig. \ref{fig:distortions}.

\noindent\textbf{Spatial Transformations}.
It is widely acknowledged that most popular CNN-based models are vulnerable to even mild spatial transformations \cite{azulay2018deep}. While data augmentation tricks can be applied, they typically yield only marginal improvements and struggle to generalize to unseen transformations. Here we investigate model robustness against 1) translation and 2) rotation. Specifically, the maximum horizontal and/or vertical translation is set to $10\%$ of the image resolution, divided into eight uniform levels. The range of rotation is set to $\left [ -30\degree, 30\degree \right ]$ according to the position and orientation of the six myocardium regions, also divided into eight levels. We exclude complex nonlinear spatial transformations as they may not be label-preserving.

\noindent\textbf{Image Distortions}. We consider five types of image distortions: 1) Gaussian noise, 2) impulsive noise, 3) Rician noise, 4) Gaussian blur, and 5) JPEG compression. Gaussian noise and blur are the most frequently studied distortions. Impulsive noise is typically used to simulate artifacts that appear during certain time slots under adverse working scenarios and equipment conditions \cite{Vaseghi1996Impulsive}. Spatially homogeneous Rician noise \cite{Coupe2010Robust} arises because CMR images are often acquired using a single-channel radiofrequency coil. JPEG compression is also commonly applied before storing CMR images. 
We set five intensity levels for each type of distortion.

\noindent\textbf{Adversarial Attacks} generate visual stimuli that appear almost identical to the original examples but can lead to catastrophic failures in CNNs  \cite{Goodfellow2014Explaining, Madry2017Towards}. 
The integration of SPT serves as a defense mechanism against such attacks. To evaluate its adversarial robustness, we employ projected gradient descent without constraints \cite{Madry2017Towards} to create adversarial examples. The step size and iteration number together determine the intensity levels. For an $8$-bit image, the step size is set to $\{1/255, 2/255, 4/255, 8/255, 16/255, 24/255, 32/255, 48/255\}$, and the number of iterations is set to $\{50, 100\}$.

\section{Experiments}\label{sec:experiment}
In this section, we describe the experimental setup, present the prediction accuracy results, and most importantly offer a thorough assessment of our model robustness against input perturbations. Additionally, we conduct a series of ablation experiments to confirm that the integration of SPT indeed enables robust LV quantification.

\subsection{Experimental Setup}\label{subsec:setup}
\noindent\textbf{Dataset}. We use the Cardiac-DIG dataset~\cite{xue2021left}, which comprises CMR data from $145$ patients with diverse pathologies. Each patient contributes to a sequence of short-axis CMR slices, which includes $20$ mid-cavity slices, leading to a total of $145\times 20 = 2,900$ images. Each image is annotated with eleven indices, including the cavity and myocardium areas, three directional dimensions, and six regional wall thicknesses.

\noindent\textbf{Training and Evaluation Procedure}. Training is carried out by minimizing $\ell(\bm \theta, {\bm \Sigma}_\mathrm{lr})$ with $\lambda=10^{-3}$ using Adam optimizer. The initial learning rate is $1\times10^{-3}$ and is decayed by multiplying $0.2$ when the loss plateaus. The minibatch size is $80$, containing CMR data from four subjects. The maximum epoch number is $400$. We apply standard data augmentation tricks that are widely used in training CNN-based methods for LV quantification, including random cropping, translation, rotation, and contrast adjustment. 

We adopt the MAE and Pearson correlation coefficient $\rho$ to evaluate the prediction accuracy and robustness. Five-fold cross-validation is implemented, where we divide the whole dataset equally into five groups, each with $29$ subjects. Four groups of samples are used for training ($70\%$, $101$ subjects) and validation ($10\%$, $15$ subjects), and the last group is used for testing ($20\%$, $29$ subjects). The mean and standard deviation results are reported.

\begin{table*}[t] 
\scriptsize
\caption{Accuracy results in terms of MAE (area in $\mathrm{mm}^2$, while dimension and thickness in $\mathrm{mm}$) and Pearson correlation.}
\hspace{-10pt}\begin{tabular}{c|c|cccccccc}
\toprule
\multicolumn{2}{c|}{Method} & \makecell[c]{DCAE \\ \cite{xue2017direct}} & \makecell[c]{DMTRL \\ \cite{xue2018full}} & \makecell[c]{Jang18 \\ \cite{Jang2018Full}} & \makecell[c]{Yang18 \\ \cite{yang2018left}}  & \makecell[c]{Xue20-MD \\ \cite{Xue2020Left}} & \makecell[c]{Yu21 \\ \cite{yu2021multitask}} &  \makecell[c]{Li23 \\ \cite{li2023task}} & Ours \\ \hline

\multirow{3}{*}{\makecell*[c]{\vspace{2mm} \quad \\ Area}}
& $A_{1}$  & \makecell[c]{223±193\\0.853}  &  \makecell[c]{189±159\\ 0.947} & \makecell[c]{207±174\\0.874} & \makecell[c]{219±168\\-} & \makecell[c]{199±160\\-} & \makecell[c]{176±189\\-} & \makecell[c]{\textbf{105±90}\\ \textbf{0.985}} & \makecell[c]{207±182\\0.863} \\ 

& $A_{2}$  & \makecell[c]{185±162\\0.953} & \makecell[c]{172±148\\0.943} & \makecell[c]{170±147\\0.956} & \makecell[c]{175±126\\-} & \makecell[c]{\textbf{139±111}\\-} & \makecell[c]{160±158\\-} & \makecell[c]{158±130 \\ {0.938}} & \makecell[c]{149±122 \\ \textbf{0.969}}  \\ 

& Avg & \makecell[c]{204±133\\0.903} & \makecell[c]{180±118\\0.945} & \makecell[c]{188±162\\-} & \makecell[c]{197±147\\0.935} & \makecell[c]{169±105\\-} & \makecell[c]{168±130\\-} & \makecell[c]{\textbf{132±110} \\ \textbf{0.962}} & \makecell[c]{178±114 \\ 0.955}  \\  \hline 

\multirow{4}{*}{\makecell*[c]{\vspace{2mm}\quad \\ \quad \\ Dimension}}
& $D_{1}$ & - &  \makecell[c]{2.47±1.95\\0.957} & \makecell[c]{2.55±2.08\\0.937} & \makecell[c]{2.43±1.94\\-} & \makecell[c]{2.01±1.59\\-} & \makecell[c]{2.35±1.87\\-} & \makecell[c]{\textbf{1.80±1.52}\\ \textbf{0.973}} & \makecell[c]{2.01±1.66\\ 0.946} \\ 

& $D_{2}$ & - & \makecell[c]{2.59±2.07\\0.854} & \makecell[c]{2.16±1.85\\ 0.958} & \makecell[c]{2.52±1.99\\-} & \makecell[c]{2.38±1.62\\-} & \makecell[c]{2.27±2.32\\-} & \makecell[c]{\textbf{1.79±1.49} \\ \textbf{0.977}} & \makecell[c]{2.42±1.9 \\ 0.964} \\ 

& $D_{3}$ & - & \makecell[c]{2.48±2.34\\0.943} & \makecell[c]{2.56±2.08\\0.941} & \makecell[c]{2.76±2.09\\-} & \makecell[c]{2.46±1.60\\-} & \makecell[c]{2.46±2.03\\-} & \makecell[c]{\textbf{1.75±1.38}\\ \textbf{0.973}} & \makecell[c]{2.38±1.92\\ 0.951} \\ 

& Avg & - &  \makecell[c]{2.51±1.58\\0.925} & \makecell[c]{2.42±2.01\\-} & \makecell[c]{2.57±2.00\\0.921} & \makecell[c]{2.28±1.29\\-} & \makecell[c]{2.36±1.45\\-} & \makecell[c]{\textbf{1.78±1.46}\\\textbf{0.978}} & \makecell[c]{2.27±1.38\\ 0.968} \\ \hline

\multirow{7}{*}{\makecell*[c]{\quad \\ \quad \\ \quad \\ \quad \\ \quad \\ Wall \\ Thickness} }
& $T_{1}$ & \makecell[c]{1.39±1.13\\0.824} & \makecell[c]{1.26±1.04 \\ 0.856} & \makecell[c]{1.33±1.11\\0.843} & \makecell[c]{1.55±1.11\\-} & \makecell[c]{1.31±1.01\\-} & \makecell[c]{1.21±1.11\\-} & \makecell[c]{\textbf{1.12±0.88} \\ \textbf{0.913}} & \makecell[c]{1.31±1.11 \\ 0.838} \\ 

& $T_{2}$ & \makecell[c]{1.51±1.21\\0.701} & \makecell[c]{1.40±1.10\\0.747} & \makecell[c]{1.37±1.12\\ 0.754} & \makecell[c]{1.45±1.12\\-} & \makecell[c]{1.36±1.03\\-} & \makecell[c]{\textbf{1.25±1.17}\\-} & \makecell[c]{1.28±1.02 \\ \textbf{0.851}} & \makecell[c]{1.6±1.28 \\ 0.641} \\ 

& $T_{3}$ & \makecell[c]{1.65±1.36\\0.671} & \makecell[c]{1.59±1.29\\ 0.693} & \makecell[c]{1.62±1.33\\0.668} & \makecell[c]{1.59±2.22\\-} & \makecell[c]{1.47±1.21\\-} & \makecell[c]{1.50±1.45\\-} & \makecell[c]{\textbf{1.35±1.13} \\ \textbf{0.811}} & \makecell[c]{1.58±1.25 \\ 0.591} \\ 

& $T_{4}$ & \makecell[c]{1.53±1.25\\ 0.698} & \makecell[c]{1.57±1.34\\0.659} & \makecell[c]{1.58±1.36\\0.644} & \makecell[c]{1.56±1.20\\-} & \makecell[c]{1.53±1.24\\-} & \makecell[c]{1.45±1.23\\-} & \makecell[c]{\textbf{1.34±1.11} \\ \textbf{0.820}} & \makecell[c]{1.41±1.24 \\ 0.605} \\ 

& $T_{5}$ & \makecell[c]{1.30±1.12\\ 0.781} &  \makecell[c]{1.32±1.10\\0.777} & \makecell[c]{1.33±1.19\\0.775} & \makecell[c]{1.40±0.99\\-} & \makecell[c]{1.22±1.05\\-} & \makecell[c]{1.34±1.19\\-} & \makecell[c]{\textbf{1.08±0.91} \\ \textbf{0.880}} & \makecell[c]{1.78±1.40 \\ 0.609} \\ 

& $T_{6}$ & \makecell[c]{1.28±1.0\\0.871} &  \makecell[c]{1.25±1.01\\ 0.877} & \makecell[c]{1.30±1.06\\0.87} & \makecell[c]{1.52±1.13\\-} & \makecell[c]{1.22±0.94\\-} & \makecell[c]{1.23±1.20\\-} & \makecell[c]{\textbf{1.11±0.94} \\ \textbf{0.924}} & \makecell[c]{1.44±1.21 \\ 0.824} \\ 

& Avg & \makecell[c]{1.44±0.71\\0.758} &  \makecell[c]{1.39±0.68\\0.768} & \makecell[c]{1.42±1.21\\-} & \makecell[c]{1.51±1.13\\0.805} & \makecell[c]{1.35±0.66\\-} & \makecell[c]{1.33±0.82\\-} & \makecell[c]{\textbf{1.16±0.99} \\ \textbf{0.924}} & \makecell[c]{1.52±0.74 \\ 0.851} \\  \bottomrule 
\end{tabular}
\renewcommand{\arraystretch}{1}
\label{accu_comparison}
\end{table*}

\subsection{Accuracy Results}
We compare our method with seven existing methods: DCAE \cite{xue2017direct}, DMTRL \cite{xue2018full}, Jang18  \cite{Jang2018Full}, Yang18 \cite{yang2018left}, Xue20-MD \cite{Xue2020Left}, Yu21 \cite{yu2021multitask}, and Li23~\cite{li2023task}. The MAE and Pearson correlation results are listed in Table  \ref{accu_comparison}, where we find that our model demonstrates comparable accuracy to current approaches. Xue20-MD \cite{Xue2020Left} needs to ensemble multiple Monte Carlo dropout models, which introduces considerable computation and time overhead. Meanwhile, Yu21 ~\cite{yu2021multitask} achieves higher accuracy by incorporating additional computed tomography (CT) information, and Li23~\cite{li2023task} enhances prediction accuracy through the use of 
 segmentation as auxiliary (but stronger) supervision. In contrast, our method requires minimal additional data from other modalities or with richer human annotations, yet it still achieves satisfactory performance. 

\begin{figure*}[t!]
\begin{minipage}[c]{1.05\textwidth}
\hspace{-6pt}\includegraphics[width=4.cm]{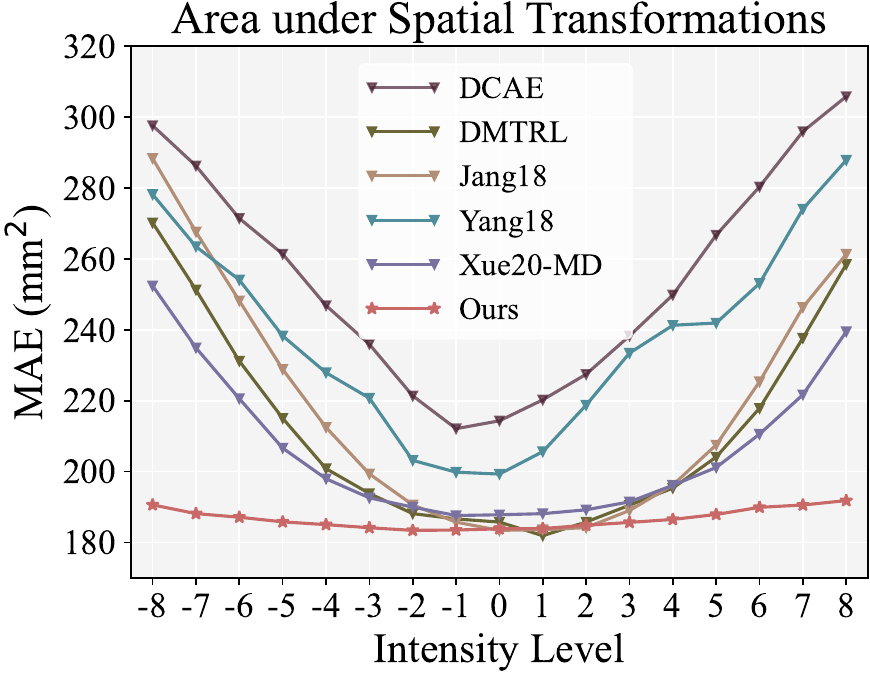}\hspace{5pt}
\includegraphics[width=4.cm]{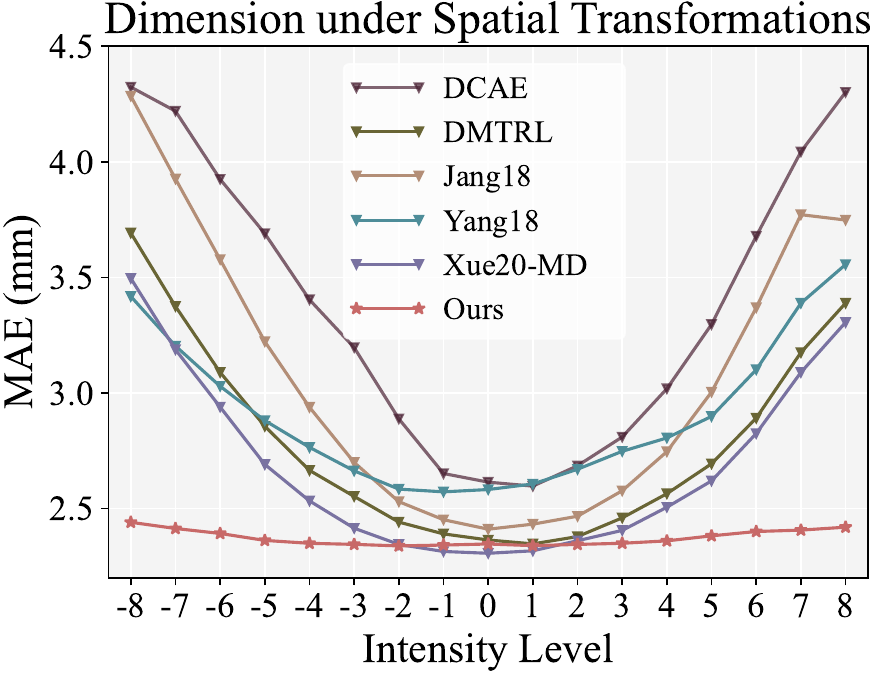}\hspace{5pt}
\includegraphics[width=4.cm]{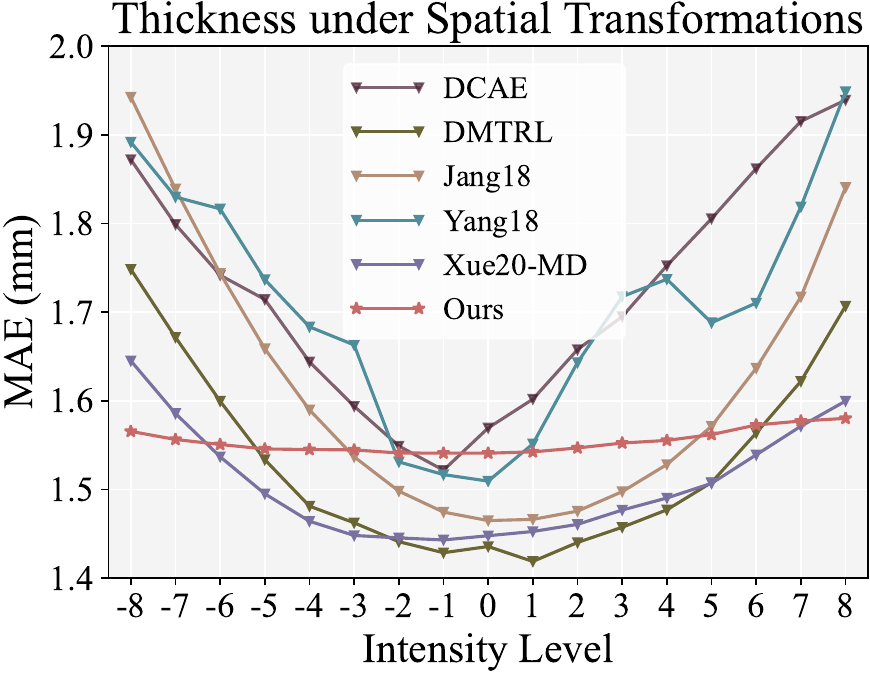}
\end{minipage} \\

\begin{minipage}[c]{1.05\textwidth}
\hspace{-6pt}\includegraphics[width=4.cm]{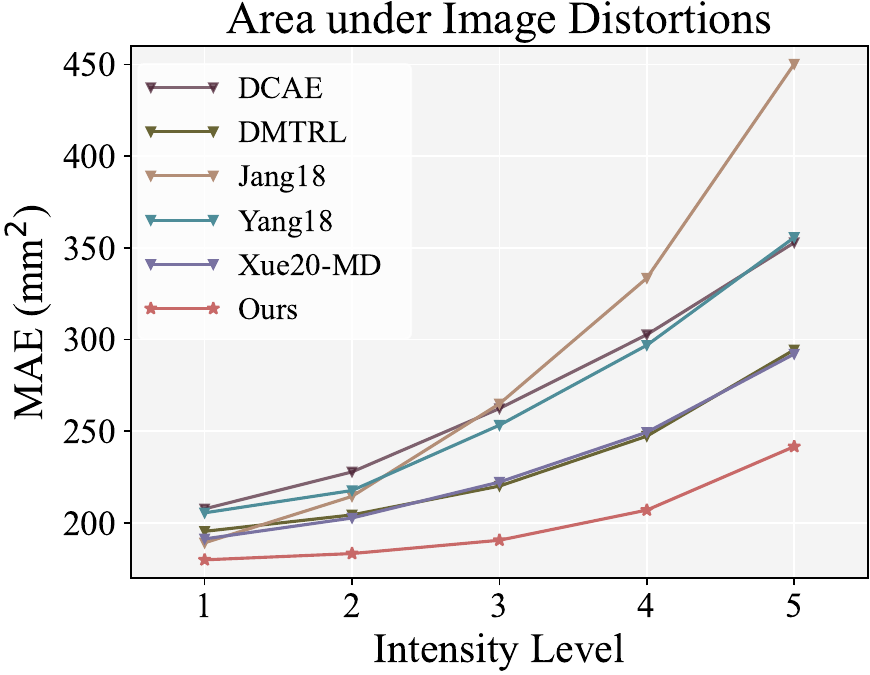}\hspace{5pt}
\includegraphics[width=4.cm]{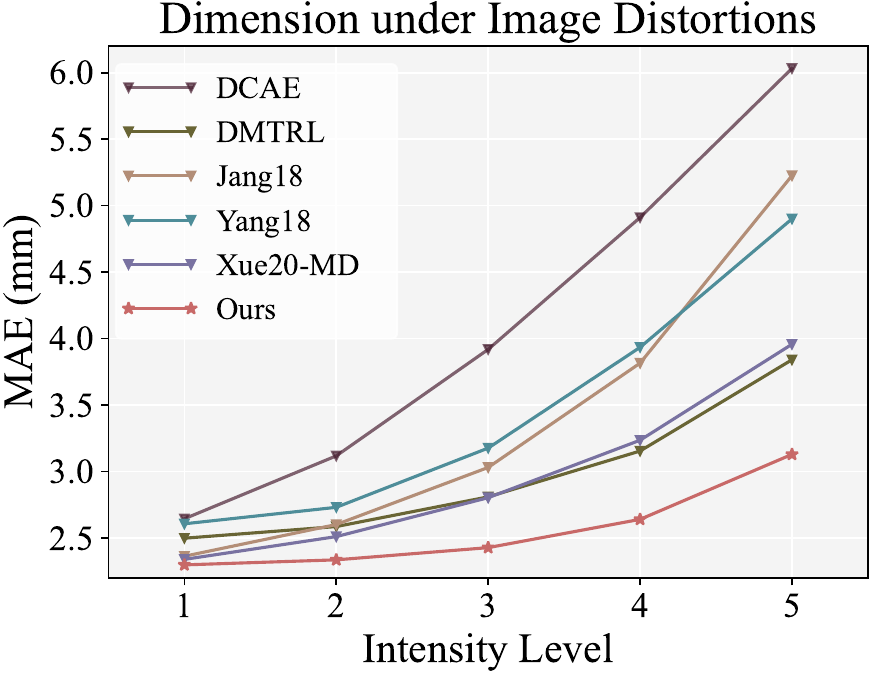}\hspace{5pt}
\includegraphics[width=4.cm]{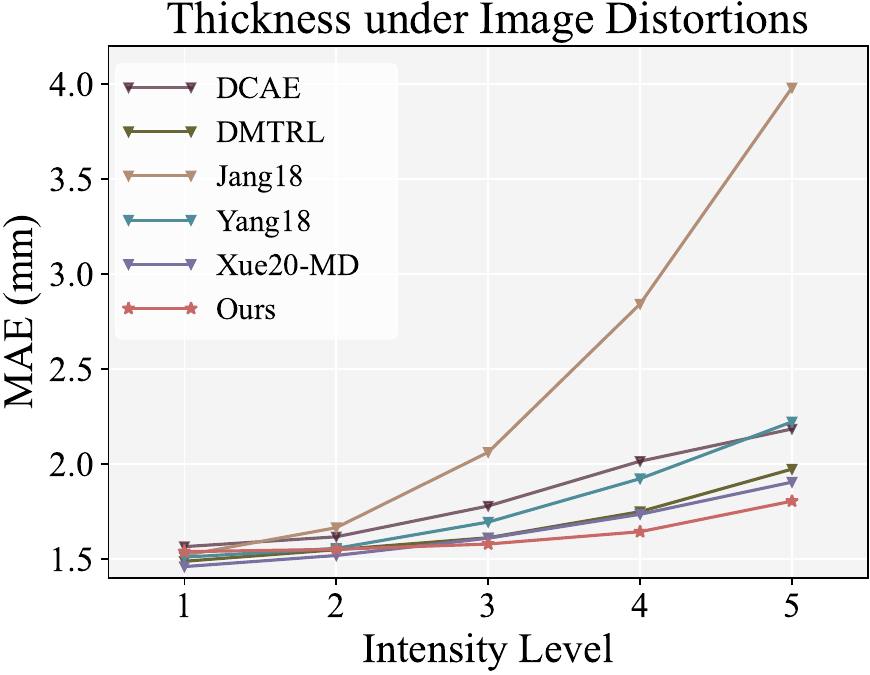}
\end{minipage} \\

\begin{minipage}[c]{1.05\textwidth}
\hspace{-6pt}\includegraphics[width=4.cm]{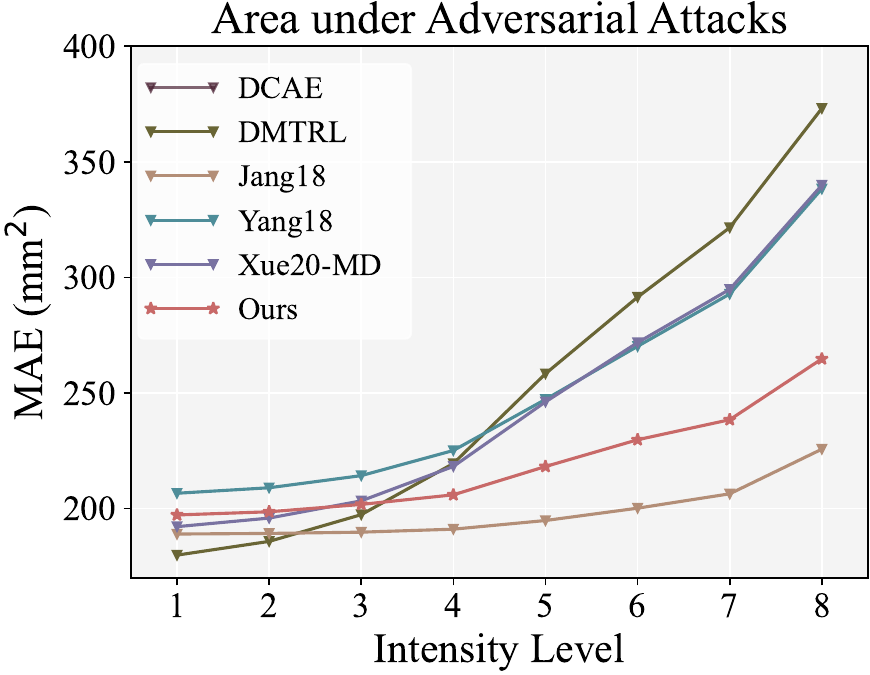}\hspace{5pt}
\includegraphics[width=4.cm]{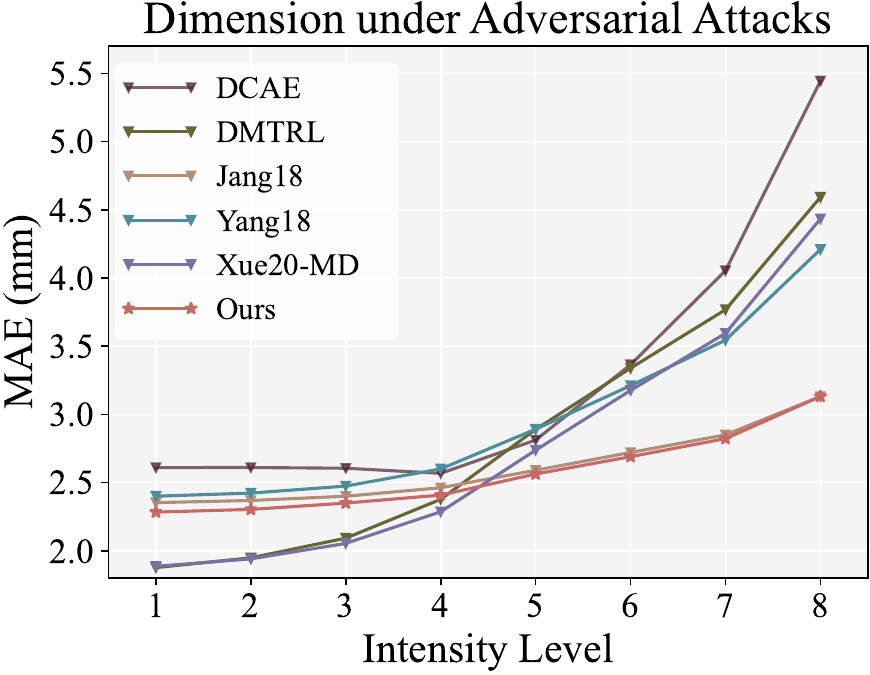}\hspace{5pt}
\includegraphics[width=4.cm]{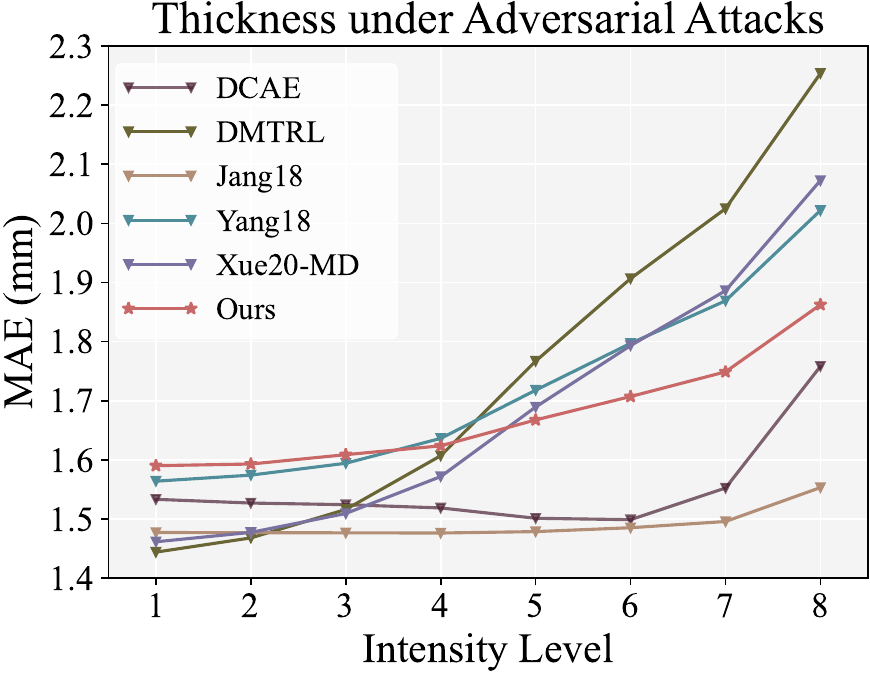}
\end{minipage} \\
\caption{Mean robustness results across three types of input perturbations. Negative intensity levels indicate translation or rotation in the opposite direction.}
\label{fig:robust_avg}
\end{figure*}

\subsection{Robustness Results}
We show the robustness results averaged across different spatial transformations, image distortions, and adversarial attacks in Fig.~\ref{fig:robust_avg}, which yields several interesting observations. First, the SPT-augmented method shows significantly improved robustness. For example, despite being blind to the five image distortions, the proposed method still achieves reliable LV quantification. Second, our method is more resilient against spatial transformations, which may be due to the translation- and rotation-equivariant properties of SPT. Third, although we do not employ adversarial training, the resulting method shows reasonable resistance to adversarial attacks.

\subsection{Ablation Studies} \label{sec:as_spt}
We design a thorough set of ablation experiments to verify that the main source of robustness is indeed from SPT. Due to the space limit, we only present the mean MAE of two areas under spatial transformations and image distortions.

We train several degenerated variants of the proposed model, in which we intentionally exclude LSTM for temporal modeling and correlation constraints for regularization. Unless otherwise specified, all variants are trained without data augmentation. Detailed specifications and acronyms of these variants are given in Table~\ref{table:variants}. \textbf{Baseline} is a generic CNN that takes a CMR slice as input and directly regresses the eleven indices. \textbf{Baseline-Aug} is built on top of Baseline by incorporating data augmentation described in Sec.~\ref{subsec:setup}. \textbf{SPT-AC} decomposes the CMR slice into two scales and three orientations,  all organized in the channel dimension. \textbf{SPT-AC-H} and \textbf{SPT-AC-L} extend SPT-AC by adding the residual highpass and lowpass subbands into the channel dimension, respectively. \textbf{SPT-OC-L} differs from SPT-AC-L by organizing the orientation subbands (along with the residual lowpass subband) in the channel dimension and the scale subbands in the batch dimension, enabling parameter sharing across scales. \textbf{SPT-SC-L} is a spatial degenerate of the proposed method, which puts the scale subbands (along with the residual lowpass subband) in the channel dimension and the orientation subbands in the batch dimension, allowing parameter sharing across orientations.

\begin{table*}[t]
\centering
\scriptsize
\caption{Variants of the proposed method. ``H'' and  ``L'' denote residual highpass and lowpass subbands, respectively.
The input shape is in the form of ``batch$ \times$ channel $\times$ height $\times$ Width''.}
\begin{tabular}{l|c|c|c|c|r|c}
\toprule
Variant  & \makecell[c]{SPT} & \makecell[c]{SPT Scales \& \\ Orientations} & \makecell[c]{H \& L} & \makecell[c]{Data \\ Augmentation}  & \makecell[c]{Input Shape} & \makecell[c]{Output Shape} \\ \hline

Baseline & \xmark & N.A. & N.A. & \xmark & $1\times 80 \times 80$ & $11 \times 1$   \\
Baseline-Aug~ & \xmark & N.A. & N.A. & \cmark & $1 \times 80 \times 80 $ & $11\times1$ \\ 

SPT-AC& \cmark & 2 / 3 & \xmark & \xmark & $6\times80 \times 80$ & $11\times1$ \\
SPT-AC-H & \cmark & 2 / 3  & H & \xmark & $7\times 80 \times 80$ & $11\times1$ \\ 
SPT-AC-L &  \cmark & 2 / 3 & L & \xmark & $7\times 80 \times 80$ & $11\times1$  \\
SPT-OC-L & \cmark & 2 / 3 & L & \xmark & $3\times 3\times 80 \times 80 $ & $11\times3$ \\ 
SPT-SC-L & \cmark & 2 / 3 & L & \xmark & $3\times 3\times 80 \times 80$ & $5\times3$\\
\bottomrule
\end{tabular}
\label{table:variants}
\end{table*}

\begin{figure*}[t]
\begin{minipage}[c]{0.49\textwidth}
\centering
\includegraphics[width=0.47\textwidth]{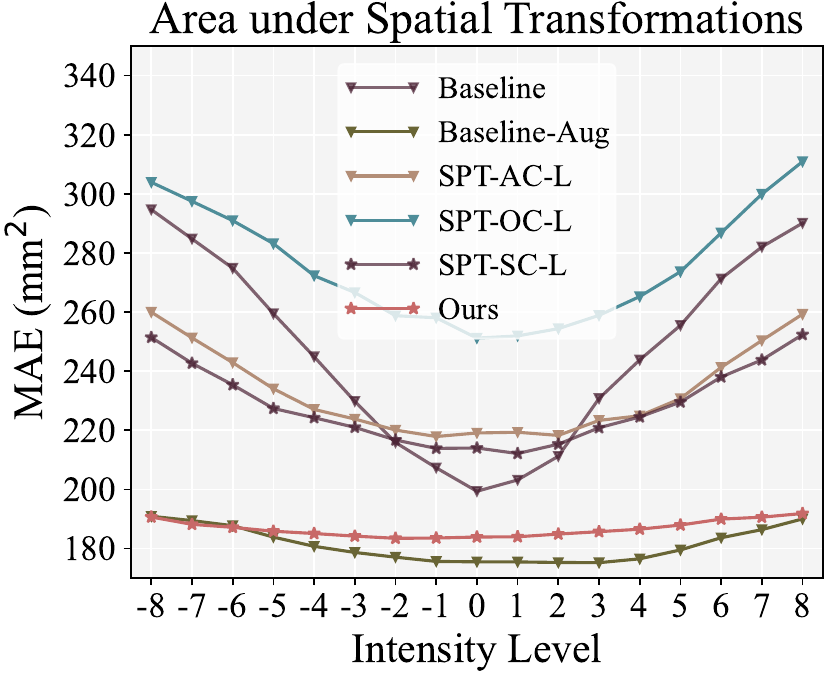}
\includegraphics[width=0.47\textwidth]{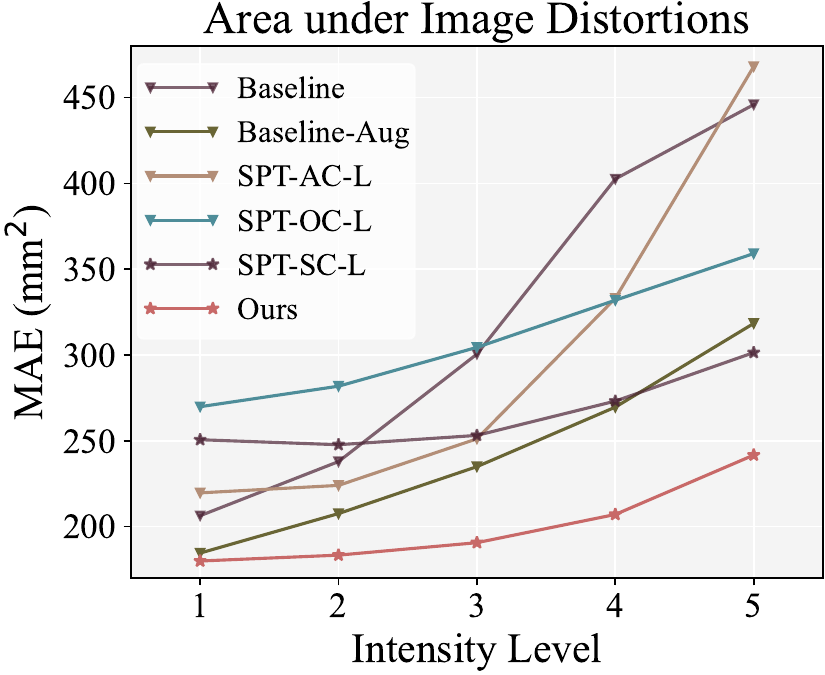} \hspace{2pt}
\caption{Identification of the role of parameter sharing in SPT.}
\label{fig:ablation_study_1}
\end{minipage}\hspace{5pt}
\begin{minipage}[c]{0.49\textwidth}
\centering
\includegraphics[width=0.47\textwidth]{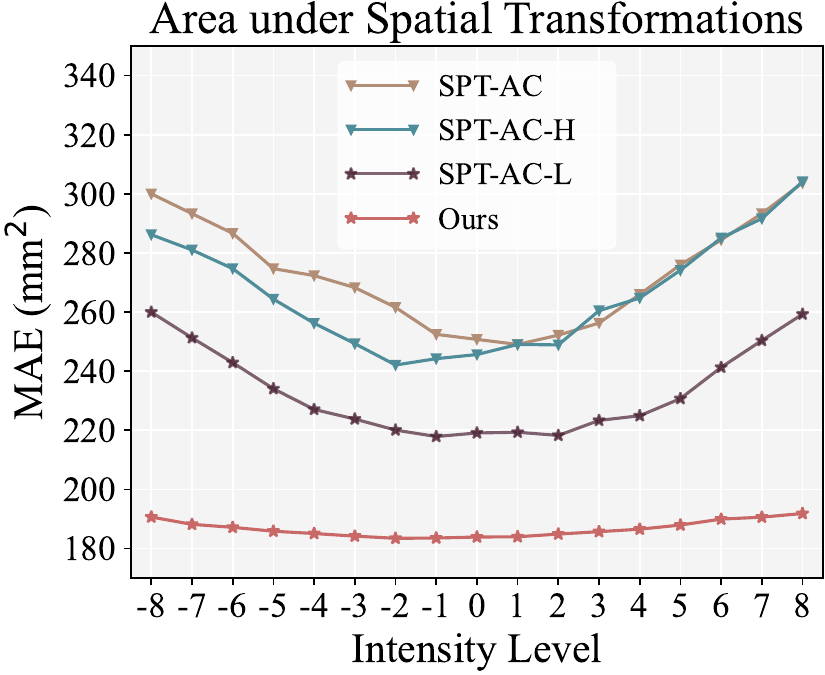}
\includegraphics[width=0.47\textwidth]{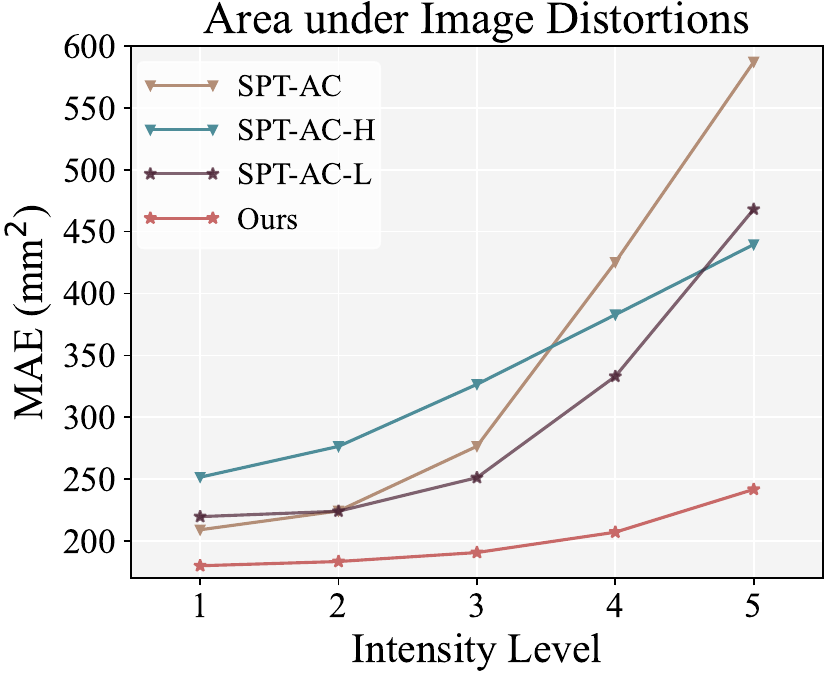} \hspace{2pt}
\caption{Identification of the role of highpass and lowpass subbands in SPT.}
\label{fig:ablation_study_2}
\end{minipage}
\end{figure*}

Through the comparison of Baseline, Baseline-Aug, SPT-AC-L, SPT-OC-L, and SPT-SC-L in Fig. \ref{fig:ablation_study_1}, we find that incorporating SPT generally enables the robustness of LV quantification.
Second, the way of sharing subbands is crucial for robustness. The variant that allows parameter sharing across orientations performs well under input perturbations. Third, compared to Baseline-Aug, SPT-augmented variants also boost model robustness under spatial transformations. 

The comparison of SPT-AC, SPT-AC-H, and SPT-AC-L in Fig. \ref{fig:ablation_study_2} highlights the impact of residual highpass and lowpass subbands on prediction robustness. Our findings show that adding the lowpass subband generally improves the robustness against input perturbations. In contrast, merely eliminating the highpass subband does not result in consistent robustness improvements. Compared to the complete model (\ie, ``Ours''), we conclude that it is the multi-scale and multi-orientation representations, combined with a proper choice of parameter sharing, that enable robustness.

\section{Conclusion}\label{sec:con}
We have presented a robust LV quantification method by incorporating SPT for fixed front-end processing. Our method yields accurate LV quantification and demonstrates robustness to various input perturbations. Our work highlights the importance of the robustness in computer-aided diagnosis, which has significant practical implications. We believe that SPT is also beneficial for other medical applications that seek scale- and orientation-dependent representations, a potential avenue for future exploration.

\begingroup
\newif\ifgobblecomma
\gobblecommafalse
\edef\FZ{?}
\edef\KM{,}
\catcode`?=\active
\catcode`,=\active
\def?{\FZ\gobblecommatrue}
\def,{\ifgobblecomma\gobblecommafalse\else\KM\fi}
\bibliographystyle{ieeetr}
\bibliography{main}
\endgroup

\end{document}